\documentstyle[12pt]{article}

\begin{document}
 \begin{center}
{\large\bf Spontaneous Symmetry Breaking, Off-diagonal Long-range Order,
and Nucleation of Quantum State}

Yu Shi\footnote{Electronic address: shiyu@alon.cc.biu.ac.il}
\\Department of Physics, Bar-Ilan University, Ramat-Gan 52900, Israel
\end{center}
\vspace{1cm}


short title: spontaneous symmetry breaking
\newpage
\begin{abstract}
Spontaneous symmetry breaking originats in quantum mechanical 
measurement of the relevant observable defining the physical 
situation, order parameter is the average of this observable. 
A modification is made on the random-phase postulate validating 
the ensemble description. Off-diagonal long-range order,
macroscopic wavefunction and interference effects in
many-particle systems present when there is a so-called
nucleation of quantum state, which is proposed to be
the origin of spontaneous gauge symmetry breaking, for which 
nonconservation of particle number $N$ is not essential. The 
approach based on nonvanishing expectation of the field operator,
 $<\hat{\psi}(\vec{r})>$, is only a coherent-state 
approximation in thermodynamic limit. When $N\,\rightarrow\,\infty$,
this approach is equivalent, but  $<\hat{\psi}(\vec{r})>$ is
not the macroscopic wavefunction.
\end{abstract}
\newpage
\section*{1. Introduction}

As a crucial concept of modern physics, spontaneous symmetry breaking
(SSB) referrs to that the ground state or vacuum does not possess the
symmetry of the Hamiltonian. This was explained by the fact that the 
near-degenearcy of the ground state and the nearby excited states makes 
the symmetry-breaking state more stable than symmetric states against
perturbation in case the physical ground state or vacuum is {\em not} the
eigenstate of the Hamiltonian \cite{exp}\cite{peirls}\cite{weinberg}. There is
another
case in which the physical ground state is an eigenstate of the Hamiltonian,
therefore furnishes a representative of the corresponding symmetry, 
the typical example is the often-quoted ferromagnetism \cite{anderson1}.
To understand why SSB occurs in this case, it seems that one usually also
have to resort to a perturbation, e.g. the auxiliary magnetic field 
in ferromagnet, which approaches zero after the thermodynamic limit is 
taken. However, there can be no external field  in reality. As indicated in this
article, the thermodynamic limit is either not essential for SSB though 
it makes SSB exact in case the physical
ground state is not an eigenstate of the Hamiltonian.
In this article we present a general  understanding of the mechanism of 
SSB based on basic principles of quantum mechanics, the word ``spontaneous''
is given a special meaning related to quantum mechanical measurement,
where there is intrinsic randomness, 
the resort to perturbation is not needed. To our opinion, perturbation-induced
symmetry breaking had better not be said to be spontaneous, though this is
a semantic problem.

More discussions will be made on off-diagonal long-range order 
(ODLRO) and spontaneous 
gauge symmetry breaking (SGSB) . A motivation is
the recent interests
in Bose-Einstein condensation (BEC) renewed by the experimental realization
using   trapped atoms \cite{atom}.  Interference 
of two condensates was discussed in 
case the  number of atoms are conserved \cite{juv}. 
This raised
some puzzles \cite{euro}
 regarding the previous  approach to macroscopic wavefunction and
 spontaneous gauge 
symmetry breaking based on $<\psi(\vec{r})>\,\neq\,0$ with
 the particle number $N$ 
 nonconserved 
\cite{goldstone}\cite{anderson}\cite{leggett}\cite{huang}, here $\hat{\psi}(\vec{r})$
is the field operator. We stress that   $<\psi(\vec{r})>\,\neq\,0$ 
is only an approximation, herein 
dubbed as ``coherent state approximation'' (CSA),
which  is a powerful approach and become exact when $N\,\rightarrow\,\infty$.
 The conservation of particle number is a presupposition for any
isolated nonrelativistic system. When $N$ is limited, 
e.g. in the BEC of ``countable''
atoms, CSA becomes not good.
We show in this article that ODLRO,
 macroscopic wavefunction and
SGSB emerge when there is a so-called nucleation of qantum state (NQS).
SGSB occurs when the phase of the wavefunction corresponding to NQS is
(relatively) determined after measurement.
These notions themselves do not necessitate  $<\psi(\vec{r})>\,\neq\,0$.
NQS directly leads to two-fluid model. Within our discussion, we introduce
a new method not using  the conventional ensemble description based on eigenstates of
the Hamiltonian. For the interference between many-particle states to be 
possible, modification should be made
on the random-phase postulate, which is the basis of 
quantum statistical mechanics. In this way, statistical mechanics is but
the quantum mechancs simplified in the case of many-particle systems
with no additional hypothesis.
We  rigorously justify the
 equivalence between
 ODLRO correponding to  a general local operator 
  and the nonvanishingness
of the coherent-state average of this operator when $N\,\rightarrow\,\infty$,
but it is shown that $<\psi(\vec{r})>$ is  {\em not} the macroscopic 
wavefunction. The possibility of the coexistence of more than one
ODLRO is also discussed from the perspective of NQS,
as well as in CSA.

\section*{2. Spontaneous symmetry breaking}
In this section, we explicitly give the general mechanism of SSB based
on basic principles of quantum mechanics, especially, it is claimed that
quantum mechanical measurement is the origin of SSB. There may be overlap
or equivalence between some points here
and previously known results, but
all the essentials are given for completeness.
   
As a foundation of quantum theory, the expansion postulate states that
any state of a system is 
  a superposition of a  set of eigenstates of an arbitrary observable
  made on it,
  \begin{equation}
 |\Psi(t)>\,=\,\sum_{j}\Phi_{j}(t)|j>, \label{eq:wave}
 \end{equation}
  where  
  $<j|k>\,=\,\delta_{jk}$, 
  $\Phi_{j}$ is the wavefunction corresponding to
 $|j>$.  Schr\"{o}dinger equation 
 $i\hbar\partial_{t}|\Psi(t)>\,=\,\hat{H}|\psi(t)>$ reduces to 
 $i\hbar \partial_{t}\Phi_{j}(t)
\,=\,\sum_{k}\Phi_{k}(t)<j|\hat{H}|k>$.
Another basic postulate states that, after measurement,
the state of the system projects, or say collapses to an eigenstate
of the relevant observable, denoted as $\hat{R}$ here. 
To which eigenstate the state collapses is intrinsicly random though
the probability is known.
The physical situation is just {\em defined} by  
this relevant observable rather than by the Hamiltonian, though the latter 
governs the evolution of the state. 
So the physical ground state, i.e. the state with least energy which can
be reached in the given situation, must be an  
eigenstate of the relevent observable $\hat{R}$,
hence may not be the ground
state of the Hamiltonian $\hat{H}$. 
Choose  $|j>$ in (\ref{eq:wave})  be the eigenstate of $\hat{R}$.
  If  $\hat{R}$ commutes  
$\hat{H}$, they have common eigenstates,  $|j>$ is also
the 
eigenstate of $\hat{H}$, then 
$<j|\hat{H}|k>\,=\,E_{j}\delta_{jk}$, $\Phi_{j}(t)\,=\,a_{j}e^{-iE_{j}t}
/\hbar$, where $a_{j}$ is 
independent of time.    
 If at time $t\,=\,0$, the state is in an eigenstate $|I>$, 
 then $|\Psi(t)>\,=\,e^{-iE_{I}t}|I>$,
i.e. the system is always in $|I>$.
Therefore after measurement  the state of system must be
 trapped 
in a stationary state and the physical ground state is just the
ground state of $\hat{H}$. If there is degeneracy, i.e. 
there are more than one stationary state
corresponding to an eigenvalue of $\hat{R}$ and $\hat{H}$,  these states form a 
(non-identical) representation of the relevant group, 
the above mechanism just 
gives rise to  one type of SSB, herein referred to as type-1 SSB. 
It may be seen that it is just the symmetry of $\hat{H}$ that keeps the 
spontaneous breaking of this  symmetry.
 
 If $\hat{R}$ is not commutative with $\hat{H}$, then   generally
 $<j|\hat{H}|k>\,\neq\,0$ even for $k\,\neq\,j$, therefore 
  $\Phi_{j}(t)\,\neq\,0$ for any $j$ even if
  $\Phi_{j}(t=0)\,=\,\delta_{jI}$. This means that 
   the state of the system cannot
  be trapped in an eigenstate of $\hat{R}$ though it is an
eigenstate of $\hat{R}$ just after measurement. 
What is interesting is that the
 probability $|\Phi_{j\neq I}(t)|^{2}$ may be very small within
  the observational 
 time. Suppose the unitary transformation from   the eigenstate of $\hat{H}$,
 say $|j'>$,
 to that of $\hat{R}$, say $|j>$, 
   is 
 $|j>\,=\,\sum_{j'}U_{jj'}|j'>$. If $|\Psi(t=0)>\,=\,|j>$, then 
 $|\Psi(t)>\,=\,\sum_{j'}U_{jj'}e^{-iE_{j'}t/\hbar}|j'>$, the
  probability for the state to be in the eigenstate $|k>$ of
  $\hat{R}$ is then 
 $|<k|\hat{\Psi}(t)>|^{2}\,=\,
 |\sum_{j'}U^{*}_{kj'}U_{jj'}e^{-iE_{j'}t/\hbar}|^{2}$,
 which is near to $\delta_{kj}$ if $E_{j'}$ is near to each other.
 Particularly, for a two-state system,
   if $|\Psi(t=0)>\,=\,|1>$, then 
 $|<2|\hat{\Psi}(t)>|^{2}\,\propto\,\sin^{2}[(E_{1}'-E_{2}')t/2\hbar]$. 
 Here $|1>$ and $|2>$ are the two eigenstates of $\hat{R}$.
 For instance, in an ammonia
 molecule $\hat{R}$ is the position operator of the nitrogen atom.
 For the massive nutrino, $\hat{R}$ corresponds the generation and
 distinguishes $\nu_{e}$, $\nu_{\mu}$ or $\nu_{\tau}$, this 
 mechanism gives rise to the neutrino oscillation, which  however,
 must be very slow.  
  So
  if the eigenstate  of $\hat{R}$ is superposed by 
  near-degenerate eigenstates of $H$, there can be an effective SSB,
  referred to as type-2 SSB. Unlike type-1, type-2 SSB is not 
 absolute, it is valid compared with the observational time and may be
 exact for infinitely large system, or say in  thermodynamic limit.  
 Now there is no common set of eigenstates of $\hat{R}$ and $\hat{H}$, the physical 
 ground state must be a superposition of the ground state and nearby
 excitated states of $\hat{H}$.
 SSB occurs whenever a measurement is performed in 
 case there is a symmetry corresponding to the same eigenvalue of $\hat{R}$
 under the condition of degeneracy or near-degeneracy, so 
 in principle it
 is not restricted to ground state. When the eigenvalue of $\hat{R}$ is
 zero, the symmetry is not broken.
    We also note that in the case without symmetry, the same 
  mechanism may lead to spontaneous 
 ergodicity breaking. 
Our classification here is consistent with  that of Peirls \cite{peirls},
but he attribute the type-two here to  perturbations while no explaination
was made on type-1, so there are essential differences.
 We expose that it is the the projection of quantum 
mechanical measurement  that originats the result that the physical state, 
usually the 
ground state, is in one of the eigenstates, which are related to each other
by symmetry,  rather than in a superposition which
possess symmetry explicitly. The degeneracy or near-degeneracy is a guarantee
of this result. In this way, as an  essential element of SSB,
 spontaneousness can be thought to be related to the 
 intrinsic quantum mechanical randomness, i.e. one cannot predict 
 to which eigenstate of $\hat{R}$ the system collapses  in a
 single run of measurement.   The 
symmetry breaking in classical systems is simply
induced by a perturbation, and there is no possibility of superposition at all, so we 
suggest 
it had better not  be said to be spontaneous.

Let us examine the prototype of  type-1 SSB, the
ferromagnetism (FM).
$\hat{H}\,=\,\sum_{(ij)}J_{ij}(S^{z}_{i}S^{z}_{j}
+S^{+}_{i}S^{-}_{j})$ with $J_{ij}\,<\,0$,
 $S_{i}^{\pm}\,=\,S_{i}^{x}\,\pm\,iS_{i}^{y}$. In this case of isotropy
 and without magnetic field,
 it should be stressed that {\em $z$
direction is only determined after measurement}. 
$\hat{R}\,=\,\sum_{i} S^{z}_{i}$. $[\hat{H},  S^{z}_{i}]\,=\,0$ and
thus $[\hat{H}, \hat{R}]\,=\,0$, so the physical ground state is just
the ground state  of $\hat{H}$, and 
 is also an eigenstate of $S^{z}_{i}$. 
 Since $S_{i}^{\pm}|S_{i}^{z}>\,=\,\sqrt{S(S+1)-S_{z}(S_{z}\pm 1)}
 |S_{i}^{z}\pm1>$,
the contribution of second term of $\hat{H}$  always
 gives zero. Thus in ground state 
all $S^{z}_{i}$ have the largest eigenvalue $S$ to make the eigenvalue
 of $\hat{H}$ lowest. The ground state is
  degenerate with  $SO(3)$ symmetry, meaning that 
 $z$ direction
  is arbitrary. But it is determined after measurement, which
   direction is 
  determined to be $z$ direction is random. Whenever it is determined,
 it is kept forever.  The usual method of
  applying an infinitesimal magnetic field is a  practical way of 
 calculation but can not serve as an explaination for SSB in quantum 
 system, though  there is no probelm for classical Ising model \cite{yang2}.
   If
  there is an easy axis, there is no symmetry to break. If there is
  an easy plane, there is an SO(2) symmetry to be spontaneously broken after
  measurement. SSB at a finite temperature  referrs to that the magnetization
  with  non-zero but
  not the  largest value spontaneously select a direction after measurement.
   Therefore resorting to the project postulate of quantum mechanical measurement,
we understand that {\em spontaneous magnetization can occur without 
external magnetic field}.
  
   Type-2 SSB is exhibited in systems with two or more alternative states
   with a non-zero transition probability between each other,
 e.g. molecules such as hydrogen, ammonia and sugar, as well as
$K$-meson and the possible massive neutrino, etc. 
For antiferromagnetism (AFM) on a bipartite lattice,
$\hat{R}\,=\,\sum_{i}S^{z}_{i}-\sum_{j}S^{z}_{j}$, where $i$, $j$ belong 
to two
sublattices, respectively. $[\hat{H},\hat{R}]\,\neq\,0$,
so SSB in AFM is type-2, the physical ground state must be a combination of
the ground state   and the nearby excited states of $\hat{H}$ . 
This is supported by the fact that 
Neel state is not an eigenstate of $\hat{H}$, and  by that there are
excitations proportional to $1/N$ \cite{and2}\cite{peirls}.   
The near-degeneracy in AFM and its difference with FM was emphasized
\cite{anderson1}.

\section*{3. Random-phase postulate and  interference between
many-particle states}

Consider interference between two states $|\Psi(a)>$ and $|\Psi(b)>$, 
one obtains from 
(\ref{eq:wave})  
that $<\Psi(a)|\Psi(b)> \,=\, \sum_{j}|\Phi_{j}(a)|^{*}|\Phi_{j}(b)|
e^{i[\phi_{j}(b)-\phi_{j}(a)]}$, where $\phi_{j}$ is the phase of $\Phi_{j}$.
 As the basis of quantum statistical mechanics,
the random phase postulate
states that the time average of $\Phi_{i}^{*}\Phi_{j}$ vanishes for
 $i\,\neq\,j$ over 
  an interval short compared to the resolving time of observation but long
   compared to molecular time, therefore the system can be described as an 
   incoherent superposition of stationary state
\cite{huang}. The average of quantity $\hat{O}$, 
$<\Psi|\hat{O}|\Psi>\,=\, \sum _{ij}\Phi_{i}^{*}\Phi_{j}<i|\hat{O}|j>$ 
consequently reduces to 
$\sum _{i}\rho_{i}<i|\hat{O}|i>$,  $\rho_{i}$ make up the density matrix,
which is diagonal. In some textbooks \cite{feynman}, the
diagonalization of density matrix is approached by arguing that there is
always a certain
representation where the density matrix is diagonal. 
This is insufficient, since it is uncertain that
this representation is 
that of Hamiltonian. Thus the
 random-phase potulate is necessary 
to make the density matrix
diagonal in every representation, therefore in that of Hamiltonian.
However, for the time average of $\Phi_{i}^{*}\Phi_{j}$ 
to vanish, each phase must randomly change with time. Therefore
the system 
   cannot be isolated \cite{huang}, but on the contrary 
   the result  is actually  applied to isolated systems. 
   Another unsatisfactory point is that an additional time average is
   made on the quantum mechanical average. But in fact  the element of
   time average has been contained in quantum mechanical average.
    Furthermore,
  {\em  this postulate implies that there cannot
    be interference among many-states systems}, since the time average
    of each term in 
 $<\Psi(a)|\Psi(b)>$ would vanish under this assumption.
Therefore the assumption  that each phase change with time  has actually
    been falsified 
by Josephson effect and will by other 
interference effects in  many-body systems such as  Bose 
condensates.
 Here we modify the random phase postulate to that {\em each phase $\phi_{j}$ is
  random and unpredictable but does not change with time and is determined
   after measurement}. Of course  only relative values are
   meaningful for phases.
 The assumption that the phases are random
at each time was made in the 
derivation of the master equation \cite{hove}, however,
to our knowledge,
there was no attempt in assuming that they do not change with time.
    The randomness and unpredictability is consistent
   with the gauge symmetry, the determination
   of the phase  is just SGSB.
This modification
is actually a supplement  to the expansion postulate,
its validity is independent of the particle number. 
Many-body effect exhibits in that 
when the number of eigenstates $\rightarrow\,\infty$, 
$\Phi_{i}^{*}\Phi_{j}<i|\hat{O}|j>$
 with $i\,\neq\,j$  cancel each
  other since the differences
 of phases are consequently also random, thus
the ensemble description is validated though generally $\Phi_{i}^{*}\Phi_{j}$
does not vanish. 
 Hence the  ensemble average or say
thermodynamic average is just  the quantum mechanical average 
reduced under the many-baody effects, hereafter we only use the term
``average''. That the 
ensemble average equals long-time average is thus not an additional 
hypothesis of
statistical mechanics, but
because they are both quantum mechanical average by definition.
Therefore statistical mechanics is nothing but a special  case of 
quantum emchanics without additional hypotheses.
 
  Normally the interference $<\Psi(a)|\Psi(b)>$ 
 vanishes when the number of eigenstates $\rightarrow\,\infty$.
But it is interesting  that 
intereference effect
emerges when there is a dominant one among 
$|\Phi_{j}(a)|
|\Phi_{j}(b)|$, i.e. if $|\Phi_{1}(a)|$ and
$|\Phi_{1}(b)|$ corresponding to an eigenstate $|1>$
are finite fractions so that $\Phi_{1}^{*}(a)\Phi_{1}(b)$ cannot
be cancelled. In this case, however, the
 simplification
of $<\Psi|\hat{O}|\Psi>$ is not affected, since
$\Phi_{i}\Phi_{j}$ here is of the same system, the
phase difference in each off-diagonal terms is random but that
in each diagonal term is zero.
 Therefore our modification is
 quite reasonable. 
			
\section*{4. Nucleation of quantum state,
 off-diagonal long-range order and spontaneous gauge
 symmetry breaking}

We have seen that to allow the possibility of
 interference between many-particle
states, there should exist an eigenstate, say $|1>$
with $\Phi_{1}\,=\,\sqrt{\alpha}e^{i\phi_{1}}$ while
$|\Phi_{j}|\,<<\,|\Phi_{1}|$ for $j\,\neq\,1$. 
$\alpha$ denotes a finite fraction hereafter.
	 This situation may be 
termed as ``nucleation of quantum state'' (NQS). The physics of NQS is nothing
beyond BEC or ODLRO in general. However, NQS can serve as a
useful notion regarding the whole system, while BEC regards constituent 
particles.
 Note that the physical situation is defined
by the relevant observable, so when we discuss actual happening
NQS, it is most convenient to choose the set of eigenstates to  be
 that of the relevant observable. Therefore we know 
 that the eigenstate should be stationary or nearly stationary 
for effects
  of NQS to exhibit.
  This is just the condition of SSB. Actually here it is gauge 
   symmetry that is spontaneously broken, reflected in that the phase 
   of the wavefunction of the eigenstate to which the state nucleates
    is random and 
determined by measurement.

 BEC occurs when there is a finite density of particles in the   
zero-momentum state. Generally
 it can be characterized by ODLRO, i.e.
 the  nonvanishingness of one-particle  reduced density matrix 
  $<\vec{r'}|\hat{\rho}_{1}|\vec{r}>\,=\,
  <\hat{\psi}^{\dag}(\vec{r'})\hat{\psi}(\vec{r})>$  \cite{penrose} \cite{yang},
   which can be factorized as \cite{yang}\cite{anderson}
   \begin{equation}
 <\hat{\psi}^{\dag}(\vec{r'})\hat{\psi}(\vec{r})>\,=\,
 \displaystyle{\sum_{n}}
 \lambda_{n}f_{n}^{*}(\vec{r'})f_{n}(\vec{r}),	 \label{eq:odlro}
 \end{equation}
 where $f_{n}(r)$ is the eigenfunction of $\hat{\rho}_{1}$ with eigenvalue 
 $\lambda_{n}$. ODLRO is equivalent to the existence of
  $\lambda_{1}\,=\,N\alpha$.
This concept is  general for 
  reduced density matrices $\hat{\rho_{n}}$ 
  and provides a unified framework for superfluidity
  and superconductivity \cite{yang}.
 It was later thought that the  main part of (\ref{eq:odlro})
 is just $<\hat{\psi}(\vec{r'})>^{*}<\hat{\psi}(\vec{r})>$, 
 a superfluid was defined
 to be with nonvanishing $<\hat{\psi}(\vec{r})>$, 
 which was claimed to be the macroscopic wavefunction and 
 indicates SGSB \cite{anderson}.
 Similar approach   was adopted in
 BCS superconductivity (SC) \cite{bcs}. 
 A usaual  way of understanding $<\hat{\psi}(\vec{r})>\,\neq\,0$  
  is to introduce an auxiliary field which is coupled to
 $\hat{\psi}(\vec{r})$ and approaches infinitesimal in the
 thermodynamic limit \cite{huang}, but this field 
 is unphysical.
 One may argue that it may serve 
  mathematically as a Lagrangian multiplier for the constraint on 
  $N$, but the latter has been taken into account by
  chemical potential.  Further, as explained above, 
  even for ferromagetism, this method is defective and
    superfluous for  SSB.
  Alternatively one may simply think that the macroscopic wavefunction of
   condensed
 particles is just $\sqrt{N_{0}/V}b_{k}$, where $N_{0}$ is number of
  condensed particles, $b_{k}$ is the corresponding
  single-particle wavefunction. 
However,  this approach
reduces the phase  to only that of the single-particle and thus
  has no physical significance \cite{leggett}.  Here
  we show that SGSB itself does not require nonconservation
  of particle number, the macroscopic wavefunction
   is derived based on NQS.

It has actually been emphasized that 
the conservation of particle number is necessary for 
BEC \cite{huang}. Viewed in our framework, the relevant observable
$\hat{R}\,=\,\hat{N_{0}}$. For SC, it is well known that BCS
ground state \cite{bcs} 
is not an eigenstate of either $\hat{H}$ or $\hat{N}$. However, since the 
electron number is conserved, BCS ground state
is only an approximation for 
convenient calculation \cite{bcs}. There is still no exact solution of
BCS Hamiltonian. 
The real physical ground state may be conjectured to be
 $N-$particle 
projection of BCS ground state, so it is 
  $|\hat{\Psi}_{G}>\,\propto\,
(\sum_{k}g_{k}\eta^{\dag}_{k})^{N/2}|0>$, where
$k$ denotes momentum, 
$\eta^{\dag}_{k}\,=\,c_{k\uparrow,-k\downarrow}$, $g_{k}$ is Cooper
pair wavefunction in momentum space, here $N$ is assumed to be even.
$|\Psi_{G}>$ is of a standard form for a state with ODLRO \cite{yang},
 and is surely an eigenstate of $\hat{H}$ since it is
that of $\hat{N}$. The excited state with $N_{G}/2$ Cooper pairs can be
given as  $|\Psi_{E}>\,\propto\, \prod_{i=1}^{N-N_{G}}c_{k_{i}}^{\dag}
(\sum_{k}g_{k}\eta^{\dag}_{k})^{N_{G}/2}|0>$, which possesses ODLRO if 
$N_{G}\,=\,N\alpha$. This state can exist for both even and odd $N$.
Now $\hat{R}\,=\,\hat{N}_{C}\,=\,\sum_{k}\eta_{k}^{\dag}\eta_{k}$, 
the number of Cooper pairs,  the state of the system 
is always its eigenstate.
One may obtain $\hat{N}_{C}|\Psi_{G}>\,=\,(N/2)|\Psi_{G}>$,
$\hat{N}_{C}|\Psi_{E}>\,=\,(N_{G}/2)|\Psi_{C}>$. 

Without need of knowledge of the exact form of $|\Psi_{G}>$ and  
$|\Psi_{E}>$, the situation of SSB is known. Generally for both BEC and
SC, $[\hat{H},\hat{R}]\,\neq\,0$, hence SSB are type-2. When
all particles are condensed in BEC (paired in SC), SSB is type-1.
Free bosons always satisfy $[\hat{H},\hat{N}_{0}]\,=\,0$, hence SSB 
would always be type-1, but since the number of particles in each momentum 
state is conserved, BEC cannot occur unless it is prepared at the 
beginning. 

Now we show  NQS gives rise to  ODLRO,
 the nonvanishing term in ODLRO  function is factorized by a
  macroscopic wavefunction for a system with conserved number of particles.

 {\em Not reduced to the conventional ensemble description using
eigenstates of $\hat{H}$}, the one-particle ODLRO function is just 
$<\Psi|\hat{\psi}^{\dag}(\vec{r'})\hat{\psi}(\vec{r})|\Psi>$. Now we 
expand $|\Psi>$
by the eigenstate of the relevant observable $\hat{R}$. Let 
 $\hat{R}\,=\,\hat{N}_{k_{0}}$,
 the number of particles with momentum $k_{0}$, 
the eigenstates  correspond to various values of $N_{k_{0}}$. Suppose
NQS occurs  in  the eigenstate $|1>$ where 
all $N$ particles occupies 
$k_{0}$ single-particle
state. Therefore the ODLRO function is 
\begin{eqnarray}
<\Psi|\hat{\psi}(\vec{r'})^{\dag} \hat{\psi}(\vec{r})|\Psi> & = &
\frac{1}{V}\sum_{kk'}
<\Psi|a_{k'}^{\dag}a_{k}|\Psi>b^{*}_{k'}(\vec{r'})b_{k}(\vec{r})
\nonumber \\
 & = & \frac{1}{V}\sum_{j}\Phi^{*}_{j}\Phi_{j}
\sum_{k}<j|a_{k}^{\dag}a_{k}|j>
b^{*}_{k}(\vec{r'})b_{k}(\vec{r})+small\, terms \nonumber\\
& = & \frac{N}{V}\Phi^{*}_{1}\Phi_{1}
b^{*}_{k_{0}}(\vec{r'})b_{k_{0}}(\vec{r})+small\, terms \nonumber\\
 & = & W(\vec{r'})^{*}W(\vec{r})+small\,terms,	 \label{eq:fac}
\end{eqnarray}
where $W(\vec{r})\,=\,\sqrt{N/V}\Phi_{1}b_{k_{0}}\,=\,
\sqrt{N\alpha/V}b_{k_{0}}(\vec{r})e^{i\phi_{1}}$.
So we obtain the nonvanishingness and factorization of ODLRO function
 to a macroscopic wavefunction $W(\vec{r})$. which is the product of the 
 wavefunction of the whole system and that of a single particle giving
 position dependence within the system. In this way we get a  
 phase $\phi_{1}$ in addition to that of the
 single-particle wavefunction.
 $\int W^{*}(\vec{r})W(\vec{r})d\vec{r}
 \,=\,N\Phi_{1}^{*}\Phi_{1}$.  In principle, BEC
  can happen for  a non-zero $k_{0}$, but the free energy
should be made to be a minimum.
Many discussions are based on a macroscopic wavefunction, 
with the  identification as
$<\hat{\psi}>$ only serve as an irrelevant 
interpretation.
Now 
it is derived from NQS.

For fermion systems,
there cannot be NQS in single-particle states constrained by Pauli principle, 
but
it can occur in two- or even-number-particle states. This is consistent 
with that the relevant observable in SC is pair operator. In this case, 
$|\Psi>$ is expanded in two-particle states. NQS referrs to that there is a 
nonvanishing probability for all particles occupy a 
 particular two-particle state. Factorization of
 ODLRO function can be obtained in a way similar to 
 (\ref{eq:fac}).
 
 It is easy to see that 
  NQS in $n$-particle states 
 corresponds
to the existence of ODLRO for $n$-particle
states, i.e. that of eigenvalues of the order $N$ for
 $n$-particle reduced density matrix \cite{yang}. 
     Various statements
  corresponging
to ODLRO \cite{yang} can be made. For example, 
the smallest $n$ for which NQS occurs gives the collection of 
$n$ particles which form a basic group exhibiting the long-range order.
NQS in  n-particle state implies 
that
in m-particle states  for $m\,\geq\,n$. If the probability of all particles occupying
 a single-particle state is $\alpha$, then the probabilty that
 all two-particle pairs
 occupy the two-particle state formed by that single-particle state is 
obviously  $\alpha^{2}$. 

NQS directly leads to two-fluid model \cite{tisza}\cite{gorter}
\cite{london}\cite{huang} and proves its universality.
The density is $<\Psi|\hat{\psi}^{\dag}(\vec{r})\hat{\psi}(\vec{r})|\Psi>$ 
while
the current is  
$2\hbar/m\cdot
 Im(<\Psi|\hat{\psi}^{\dag}(\vec{r})\vec{\nabla}\hat{\psi}(\vec{r})|\Psi>)$,
where $\hat{\psi}(\vec{r})$ is replaced by a pair of electron operator for
SC. So the density and current of the superfluid component are
$|W(\vec{r})|^{2}$ and 
$2\hbar/m\cdot Im(W^{*}\vec{\nabla}W)$, respectively. The rest terms 
describe normal-fluid 
component. Decomposition of  free energy is more complicated, as indicated
by Gorter-Casimir model \cite{gorter}.

NQS implies that {\em the precise meaning of BEC is that the probability
that all particles are in a  particular 
momentum state is a nonvanishing $\alpha$}, 
with
an objectively random phase, which however does not change with time.
 SGSB occurs when this phase is measured.
 In conventional 
 ensemble description only
 the average of $N_{k}$ is given since it is not  a conserved quantity.
 The only exception is that  $N_{k_{0}}$   is conserved
 when $N_{k_{0}}\,=\,N$, but in this case  $\alpha\,=\,1$ and thus
 the probability for all particles occupy $k_{0}$ state is unity.
So there is no contradiction between  the present 
 and the conventional  viewpoints. 
   When the probability for the state of a system to be 
in an eigenstate is unity, there is still an objectively random phase to
be determined after measurement. Of course, only relative phase is
 meaningful.
NQS is necessary for SGSB,  their  existence is
 independent of whether the particle number is conserved.
  A free particle, for example, is always in the state of NQS with
  probability unity, SSB of gauge symmetry occurs when its (relative) phase
  is measured.
NQS may possibly exist in spaces other than momentum space, it is
 interesting
to explore BEC in real space to make an ultra-dense object,
 possessing ODLRO in momentum space.

 Certainly it is possible that NQS occurs in two or more eigenstates or
 non-orthogonal states, This is usual for few-particle systems.
 For many-particle systems, this natually leads to the 
 coexistence of more than one ODLRO. This possible situation  was first 
 considered 
 in \cite{gorkov} for Cooper-paired states with nonzero angular momentum,
 and was disfavored in \cite{anderson}. Recently there
 are more interests in this topic, as for example  in \cite{zhang}\cite{shi1}
 \cite{tian}\cite{shi2}.

\section*{5. Order parameter}
 We have seen that SSB widely exists, not limited to phase transition.
 Phase transition gives rise to symmetry breaking.
  For classical system,
 it is induced by a perturbation which may be very small. For quantum system
 it is SSB. Since the physical situation is just defined by the
 relevant observable, the symmetry breaking is reflected in the change of
its value. This applies also for non-spontaneous symmetry breaking.
 Naturally, order parameter, which designates the change of symmetry,
  is just the average of $\hat{R}$,
 e.g. that of FM phase transition is the magnetization. The mass  
 of liquid can serve as that for gas-to-liquid transition, NQS is just its
 quantum correspondence, the order parameter is obviously
 the probability 
 of nucleating to the particular eigenstate. Thus {\em the order parameter
  of BEC (SC)
  is the fraction of    
 condensed (paired) particles}, and is therefore equivalent to 
 the magnitude of the macroscopic wavefunction and thus does not contradict
 Ginzburg-Landau theory, it is also proportional to the nonvanishing 
 term in ODLRO function.  Not only for phase transition, either not only for
 SSB, order parameters can  also 
 be defined
 for other symmetry breakings.
 Thermodynamic limit is necessary for phase transitions, but may be unnecessary for
other SSB though it  makes type-2 SSB exact.
\section*{6. Coherent-state approximation}

It is well known that the coherent state $|\Psi_{c}>$
is formed by states with
different number of particles $N$,  say $|\Psi_{N}>$,
\begin{equation}
 |\Psi_{c}> \,=\,\sum_{N}f_{N}|\Psi_{N}>, \label{eq:coh}
 \end{equation}
 where the coefficients $f_{N}$ and thus the weights in expectation
 values center around the mean value of $N$, say $<N>$, with a spread of
 $\sim\,\sqrt{<N>}$ \cite{anderson}\cite{bcs}\cite{peirls}.
 Therefore CSA is quite similar to the grand canonical ensemble approach.
 We have shown that coherent  state is not essential for SGSB.
 However, CSA is a convenient approach which becomes exact when
 $N\,\rightarrow\,\infty$.

 The concept of ODLRO can be generalized to other local operator, denoted
 as $\hat{B}(\vec{r})$. For a $N-$particle system , the long-range order
 of $\hat{B}(\vec{r})$ exists if and only if 
 $<\Psi_{N}|\hat{B}^{\dagger}(\vec{r'})\hat{B}(\vec{r})|\Psi_{N}>\,\rightarrow\,
 \rho\alpha$
 as $|\vec{r}-\vec{r}'|\,\rightarrow\,\infty$, 
 here $\rho$ is the density $N/V$, which is kept constant for different 
 $N$. In this section we establish the
 equivalence between ODLRO in a system with a conserved number of particles
 and the nonvanishingness of coherent-state expectation of the corresponding operator, 
 e.g. the field operator $\hat{\psi}(\vec{r})$.
 
 Consider the ODLRO function for coherent state $|\Psi_{c}>$,
\begin{eqnarray}
 <\Psi_{c}|\hat{B}^{\dagger}(\vec{r'})\hat{B}(\vec{r})|\Psi_{c}>&=&
 \sum_{N} |f_{N}|^{2}<\Psi_{N}|\hat{B}^{\dagger}(\vec{r'})\hat{B}(\vec{r})|\Psi_{N}>
 \nonumber\\
  &=&\rho\alpha
  \label{eq:cs}
  \end{eqnarray}
 In the derivation, we used the property that 
 $\hat{B}^{\dagger}(\vec{r'})\hat{B}(\vec{r})$ conserve the particle
 number. Thus there is ODLRO in $|\Psi_{c}>$ if 
 and only if there is ODLRO in $|\Psi_{N}>$.
 Then the key point is the factorization of
 $<\Psi_{c}|\hat{B}^{\dagger}(\vec{r'})\hat{B}(\vec{r})|\Psi_{c}>$, which
has been  widely taken for granted.

Note that {\em $|\Psi_{c}>$ is not the reality but a subjective 
 construction with
infinite terms}.
Therefore one can always find a set of $f_{N}$ so that $|\Psi_{c}>$ is
an eigen-vector of $\hat{B}(\vec{r})$. Furthermore, $|\Psi_{c}>$ can be
an eigenstate of {\em any} local operator since there are infinite 
$f_{N}$-s. 
Hence
\begin{eqnarray} 
<\Psi_{c}|\hat{B}^{\dagger}(\vec{r'})\hat{B}(\vec{r})|\Psi_{c}>&=&
\sum_{\beta}<\Psi_{c}|\hat{B}^{\dagger}(\vec{r'})|\beta><\beta|\hat{B}(\vec{r})|\Psi_{c}>
\nonumber\\
&=&<\Psi_{c}|\hat{B}^{\dagger}(\vec{r'})|\Psi_{c}><\Psi_{c}|\hat{B}(\vec{r})|\Psi_{c}>,
\end{eqnarray}
where \{$|\beta>$\} is the complet set of eigenstates to which $|\Psi_{c}>$
belongs.

Therefore we proved the rogorous equivalence between ODLRO and the 
nonvanishingness of the coherent-state
expectation of the corresponding operator when $N\,\rightarrow\,\infty$.
This justifies many researches concerning long-range orders in 
condensed matter physics. For instance,
if there exists the following commutation relation
\begin{equation}
[\hat{H}_{N},\,\hat{A}(\vec{r})]\,=\,\gamma_{B}B(\vec{r})+\gamma_{C}C(\vec{r}),
\label{eq:com}
\end{equation}
where  $\hat{H}_{N}$ is the Hamiltonian of $N$-particle system, 
$\hat{A}(\vec{r})$, $B(\vec{r})$ and  $C(\vec{r})$ are local operators,
 $\gamma_{B}$ and  $\gamma_{C}$ are complex constants, one may
 claim that the long-range orders of $B(\vec{r})$ and $C(\vec{r})$ must be either
 present or absent simultaneously. In particular, if there is only
 one local operator appears on the right-hand side of (\ref{eq:com}), its
 long-range order must be absent. This conclusion  can be obtained simply by
 calculating the coherent-state expectation of (\ref{eq:com}) and
 noting that of
 the left-hand side vanishes. This method was applied to obtain 
 constraints on pairings within the Hubbard model \cite{zhang}\cite{shi1}.
 However, it was argued that this approach
  is not rigorous since the eigenvalue of the Hamiltonian for an infinite
 system is ill-defined, alternative theorems were given regarding 
 $N$-particle system \cite{tian}.
 Through the above justification, it is clear that these two approaches are 
 rigrously equivalent when $N\,\rightarrow\,\infty$. 
 Especially, the property of $|\Psi_{c}>$ that it is an eigenstate of any
 local operator is crucial in obtaining the 
 simultaneous nonvanishingness  of the expectation values 
 of $B(\vec{r})$ and $C(\vec{r})$ generally.
Because  it is the coherent-state  $|\Psi_{c}>$ that of which the
expectation  is calculating for (\ref{eq:com}), 
 and $|\Psi_{c}>$ is a superposition of $|\Psi_{N}>$ with different values of $N$,
  only eigenvalue of $\hat{H}_{N}$ with different values of
 $N$ are dealt with in obtaining the coherent-state expectation of
 (\ref{eq:com}), the problem of ill-definition is
  eliminated.

It is easy to see that NQS in the state with the particle number conserved
is equivalent to NQS in the coherent state, since the expansion 
structure (\ref{eq:wave}) of $|\Psi_{N}>$ in (\ref{eq:coh})
 is the same for different values of $N$. The macroscopic
function in CSA remains the same as the original one, given $\rho\,=\,N/V$
is constant for different $N$.
 
 When $N\,\rightarrow\,\infty$, $|\Psi_{c}>\,\rightarrow\,f_{<N>}|\Psi_{<N>}>$.
Since $\hat{\psi}(\vec{r})\,=\,(1/\sqrt{V})\sum_{k}\hat{a}_{k}b_{k}(\vec{r})$,
$\hat{\psi}(\vec{r})|\Psi_{<N>}>\,=\,\sqrt{<N>/V}b_{k_{0}}\Phi_{1}
|1>_{<N>-1}+small\,terms$,
where $k_{0}$ denotes the single-particle state into
 which particles condense, $|1>_{N}$ is the state in which NQS occurs
for $N$-particle system. It is difficult to evaluate the exact form of
$<\hat{\psi}(\vec{r})>$, since $\hat{\psi}(\vec{r}$ changes the number of particles 
thus the sum of the small terms is possibly non-negligible. But it is clear that only
$\Phi_{i}^{*}\Phi_{j}$ instead of $\Phi_{i}$ appears in each term.
Compared with the macroscopic 
wavefunction $W(\vec{r})\,=\,\sqrt{\rho}\Phi_{1}b_{k_{0}}$,
 $<\hat{\psi}(\vec{r})>$ is certainly 
 {\em not} equal to the macroscopic wavefunction, though it can serve as
 a convenient characteristic of NQS and long-range order.
 
\section*{7. Summary} 
To summarize,  SSB originats in 
 quantum mechanical measurement of the relevant 
observable defining the physical situation, and may be
classified to two types according to whether this observable 
commutes the Hamiltonian. 
Order parameter may
be generally defined as
the average of this relevant observable.
To allow the possibility of  the interference between many-particle systems,
we modify the random phase postulate, which is  the
basis of quantum statistical mechanics. The phase of the wavefunction 
correponding to each eigenstate is random and unpredictable but 
  does not change with time. This statement is actually a supplement
   to expansion postulate of quantum mechanics. Thus
   statistical mechanics is but a
   special case of
quantum mechanics with no additional hypotheses, 
incoherent ensemble description is a natual consequence of
  many-body effects.
  Interference between many-particle systems,
ODLRO and macroscopic
wavefunction are derived from nucleation of quantum state. Spontaneous gauge
symmetry breaking occurs when the
phase of macroscopic wavefunction is determined (relatively) 
after measurement. 
 Since  quantum mechanical  measurement is still an open problem, whether
SSB is a dynamical process within a finite time or a 
discontinuous process is
beyond the scope of present work. Clarification of this issue is 
in no doubt  of great importance.

Nonconservation of particle number is not essential for SGSB. The 
coherent-state approach is an approximation which may become exact
when $N\,\rightarrow\,\infty$. Although the nonvanishing
$<\hat{\psi}(\vec{r})>$ under CSA can
serve as a characteristic of long-range order, it is {\em not}
the macroscopic wavefunction.
   
Theories of spontaneous gauge symmetry breaking in relativistic field theory
are based on assuming the nonvanishingness of vacuum expectation of a scalar
field. This conflicts the basic fact that the Hilbert space is
spanned by states with definite number of particles and/or 
antiparticles in each mode. We made an attempt to modify these theories
to eliminate the  assumption that the vacuum expectation of the scalar field is
nonvanishing.
\cite{shi}.

The present  work also reminds that there is no classical-quantum boundary, 
principles of quantum mechanics apply in all scales.
``Classicality'' is an emergent property of systems with
many degrees of freedom in the absence of
NQS. The macroscopic quantum effect in superconductivity
and superfluidity first proposed by London \cite{london}
is  a consequence of NQS.


\begin{thebibliography}{anderson}
\bibitem{anderson1} One may claim  this case not be  SSB, see, e.g.,
Anderson  P W 1990 {\em Phys. Today} 117, however, as commented in
\cite{peirls}, this is problem of semantics, but it would be inconvenient
not to include this important
case. Actually most literature quote ferromagnet
as the introductory example of SSB. The present article show the common origin of
this case and the case in which the physical
ground state is not an eigenstate
of the Hamiltonian
\end{thebibliography}
\end{document}